\documentclass[aps,prl,twocolumn,superscriptaddress]{revtex4}

\usepackage{graphicx,amsmath,bm,textcomp}
\usepackage{color}
\newcommand{\unit}[1]{\ensuremath{\, \mathrm{#1}}}

\begin{document}

\title{Scaling of Traction Forces with Size of Cohesive Cell Colonies}
\author{Aaron~F.~Mertz}
\affiliation{Department of Physics, Yale University, New Haven, Connecticut 06520, USA}
\author{Shiladitya~Banerjee}
\affiliation{Department of Physics, Syracuse University, Syracuse, New York 13244, USA}
\author{Yonglu~Che}
\affiliation{Department of Molecular, Cellular, and Developmental Biology, Yale University, New Haven, Connecticut 06520, USA}
\affiliation{Department of Physics, Yale University, New Haven, Connecticut 06520, USA}
\author{Guy~K.~German}
\affiliation{Department of Mechanical Engineering \& Materials Science, Yale University, New Haven, Connecticut 06520, USA}
\author{Ye~Xu~(??)}
\affiliation{Department of Mechanical Engineering \& Materials Science, Yale University, New Haven, Connecticut 06520, USA}
\author{Callen~Hyland}
\affiliation{Department of Molecular, Cellular, and Developmental Biology, Yale University, New Haven, Connecticut 06520, USA}
\author{M.~Cristina~Marchetti}
\affiliation{Department of Physics, Syracuse University, Syracuse, New York 13244, USA}
\affiliation{Syracuse Biomaterials Institute, Syracuse University, Syracuse, New York 13244, USA}
\author{Valerie~Horsley}
\affiliation{Department of Molecular, Cellular, and Developmental Biology, Yale University, New Haven, Connecticut 06520, USA}
\author{Eric~R.~Dufresne}
\email[]{eric.dufresne@yale.edu}
\affiliation{Department of Mechanical Engineering \& Materials Science, Yale University, New Haven, Connecticut 06520, USA}
\affiliation{Department of Chemical \& Environmental Engineering, Yale University, New Haven, Connecticut 06520, USA}
\affiliation{Department of Physics, Yale University, New Haven, Connecticut 06520, USA}
\affiliation{Department of Cell Biology, Yale University, New Haven, Connecticut 06520, USA}

\date{\today}

\begin{abstract}
{To understand how the mechanical properties of tissues emerge from interactions of multiple cells, we measure traction stresses of cohesive colonies of 1--27 cells adherent to soft substrates. 
We find that traction stresses are generally localized at the periphery of the colony and the total traction force scales with the colony radius.
For large colony sizes, the scaling appears to approach linear, suggesting the emergence of an apparent surface tension of order $10^{-3} \unit{N/m}$.
A simple model of the cell colony as a contractile elastic medium coupled to the substrate captures the spatial distribution of traction forces and the scaling of traction forces with the colony size.}
\bigskip


\end{abstract}

\maketitle

Tissues have well defined mechanical properties such as elastic modulus \cite{discher-2005}.
They can also have properties unique to active systems, such as the homeostatic pressure recently proposed theoretically as a factor in tumor growth \cite{basan-2009}.
While the mechanical behavior of individual cells has been a focus of inquiry for more than a decade \cite{lee-1994, ananthakrishnan-2007, vogel-2006, gardel-2008}, the collective mechanics of groups of cells has only recently become a topic of investigation \cite{nelson-2005, duroure-2005,trepat-2009, angelini-2010, khalil-2010,liu-2010,  saez-2010, tambe-2011,maruthamuthu-2011};
it is unknown how collective properties of tissues emerge from interactions of many cells.

In this Letter, we describe measurements of traction forces in colonies of cohesive epithelial cells adherent to soft substrates.
We find that the spatial distribution and magnitude of traction forces are more strongly influenced by the physical size of the colony than by the number of cells.
For large colonies, the total traction force, $\cal{F}$, that the cell colony exerts on the substrate appears to scale as the equivalent radius, $R$, of the colony.
This scaling suggests the emergence of a scale-free material property of the adherent tissue,
an apparent surface tension of order $10^{-3} \unit{N/m}$.
A simple physical model of adherent cell colonies as contractile elastic media captures this behavior.

\begin{figure}[hbt]
\includegraphics[width=.46\textwidth]{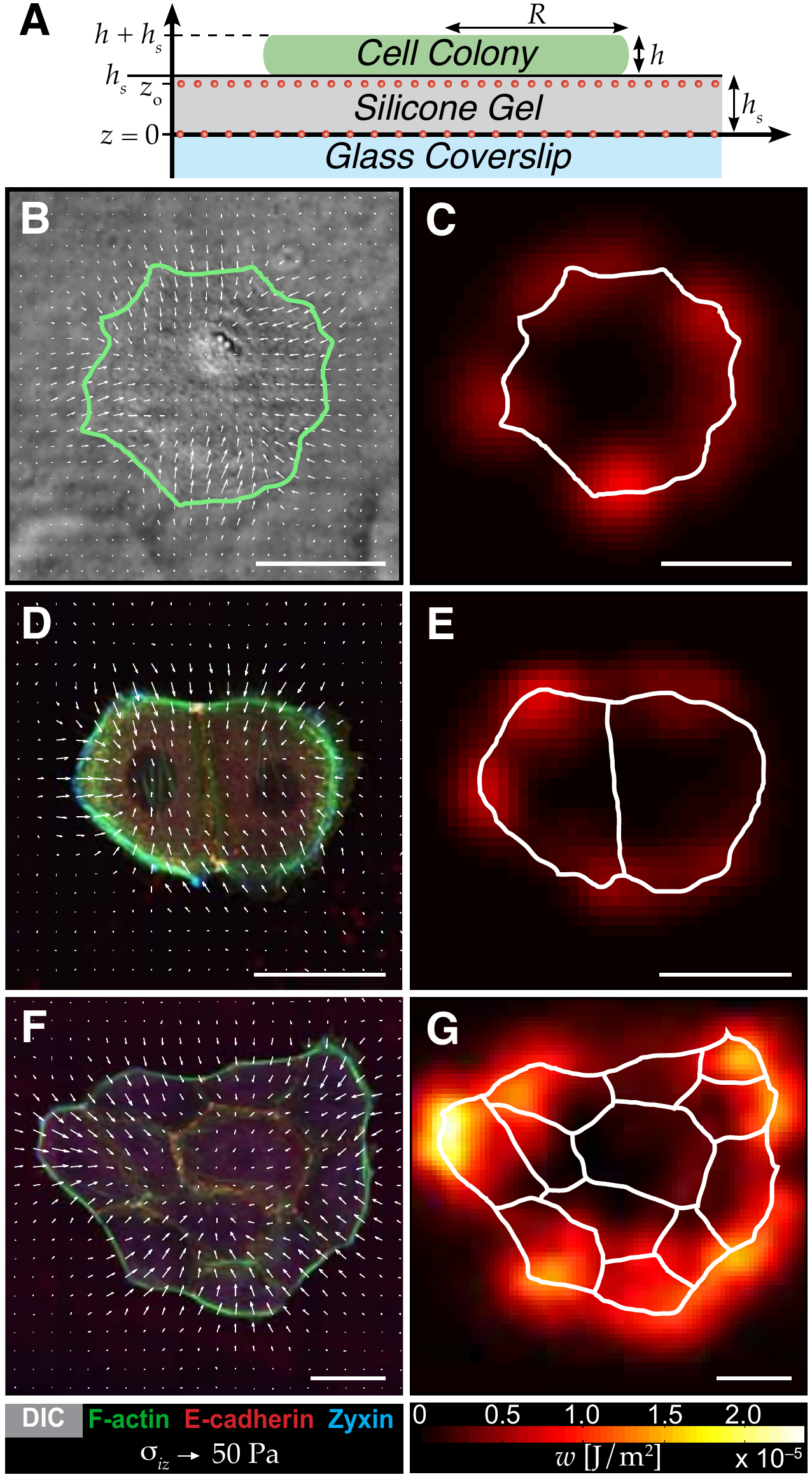}
\caption{\label{fig:TFM} (color online) Traction stresses and strain energies for colonies of cohesive keratinoctyes. 
(A) Schematic of experimental setup (not to scale) with a cell colony adherent to an elastic substrate embedded with two dilute layers of fluorescent beads.
(B, D, F) Distribution of traction stresses, $\sigma_{iz}$, and (C, E, G) strain energy, $w$, for a representative single cell, pair of cells, and colony of 12 cells.
Traction stress distribution is overlaid on a DIC image (B) or images of immunostained cells (D, F).
Solid lines in (B--C, E, G) mark cell boundaries.
For clarity, only one-quarter of the calculated stresses are shown in (B, D) and one-sixteenth of the stresses in (F).
Scale bars represent $50\unit{\mu m}$.}
\end{figure}

To measure traction stresses that cells exert on their substrate, we used traction force microscopy (TFM) \cite{dembo-1999}.
Our TFM setup consisted of a film of highly elastic silicone gel (Dow Corning Toray, CY52-276A/B) with thickness $h_s=27\unit{\mu m}$ on a rigid glass coverslip (Fig.~\ref{fig:TFM}A).
Using bulk rheology, we estimated the Young's modulus of the gel to be $3\unit{kPa}$. 
To quantify the gel deformation during our experiments, our substrates contained two dilute layers of fluorescent beads (radius $100\unit{nm}$, Invitrogen): one layer between the glass and gel and a second at height $z_{\text{o}} = 24 \unit{\mu m}$ above the coverslip \cite{xu-2010}.
To image the fluorescent beads, we used a spinning-disk confocal microscope (Andor Revolution, mounted on a Nikon Ti Eclipse inverted microscope with a 40$\times$ NA~1.3~objective).
After determining bead positions using centroid analysis in \textsc{Matlab} \cite{crocker-1996}, we calculated the substrate displacement, $u^s_i({\bf r},z_\text{o})$, across its stressed (with cells) and unstressed (with cells removed) states.
In Fourier space, the in-plane displacement field is related to the traction stresses at the surface of the substrate via linear elasticity, $\sigma^s_{iz}({\bf k},h_s) = Q_{ij}({\bf k},z_\text{o},h_s) u^s_j({\bf k},z_\text{o})$, where ${\bf k}$ represents the in-plane wave vector.
Here, $\sigma^s_{iz}({\bf k},h_s)$  and $u^s_j({\bf k},z_\text{o})$ are the Fourier transforms of the in-plane traction stress on the top surface and the displacements just below the surface, respectively.
The tensor, $Q$, depends on the thickness and modulus of the substrate, the location of the beads, and the wave vector \cite{delalamo-2007, xu-2010}.
We calculated the strain energy density, $w({\bf r})=\frac{1}{2}\sigma^s_{iz}({\bf r},h_s) u^s_i({\bf r},h_s)$, which represents the work per unit area performed by the cell colony to deform the elastic substrate \cite{butler-2001}.
The displacement of the surface was determined using $u^s_{i}({\bf k},h_s)=Q_{ij}^{-1}({\bf k},h_s,h_s) Q_{jk}({\bf k},z_{\text{o}},h_s)u^s_k({\bf k},z_{\text{o}})$.

Primary mouse keratinocytes were isolated and cultured as described in \cite{barrandon-1987}.
We plated keratinocytes on fibronectin-coated TFM substrates. 
After the cells proliferated to the desired colony sizes over 6--$72\unit{hr}$, we raised the concentration of CaCl$_2$ in the growth medium from $0.05\unit{mM}$ to $1.5\unit{mM}$. 
After 18--24$\unit{hr}$ in the high-calcium medium, cadherin-based adhesions formed between adjacent keratinocytes, which organized themselves into cohesive single-layer cell colonies \cite{okeefe-1987, vaezi-2002}.
After imaging the beads in their stressed positions, we removed the cells by applying proteinase~K and imaged the beads in their unstressed positions.

Stress fields and strain energy densities for representative colonies of one, two, and twelve keratinocytes are shown in Fig.~\ref{fig:TFM}. 
Traction stresses generically point inward, indicating that the colonies are adherent and contractile.
Regions of high strain energy appear to be localized primarily at the periphery of the single- or multi-cell colony.
For single cells, these findings are consistent with myriad previous reports on the mechanics of isolated, adherent cells \cite{wang-2001, wang-2002, delanoe-ayari-2010, fournier-2010}.
Recent reports have also observed localization of high stress at the periphery of small cell colonies on micropatterned substrates \cite{krishnan-2010} and at edges of cell monolayers \cite{nelson-2005, duroure-2005, saez-2010}.
To visualize cell--cell and cell--matrix adhesions, we immunostained multi-cell colonies for E-cadherin and zyxin.
Additionally, we stained the actin cytoskeleton with phalloidin.
Actin stress fibers were concentrated at colony peripheries and usually terminated with focal adhesions, as indicated by the presence of zyxin at the fibers' endpoints.
In contrast, E-cadherin was localized at cell--cell junctions, typically alongside small actin fibers.
Despite differences in the architecture of the relevant proteins, the stresses and strain energy distributions are remarkably similar in the single-cell and multi-cell colonies.

\begin{figure}[hbt]
\includegraphics[width=.38\textwidth]{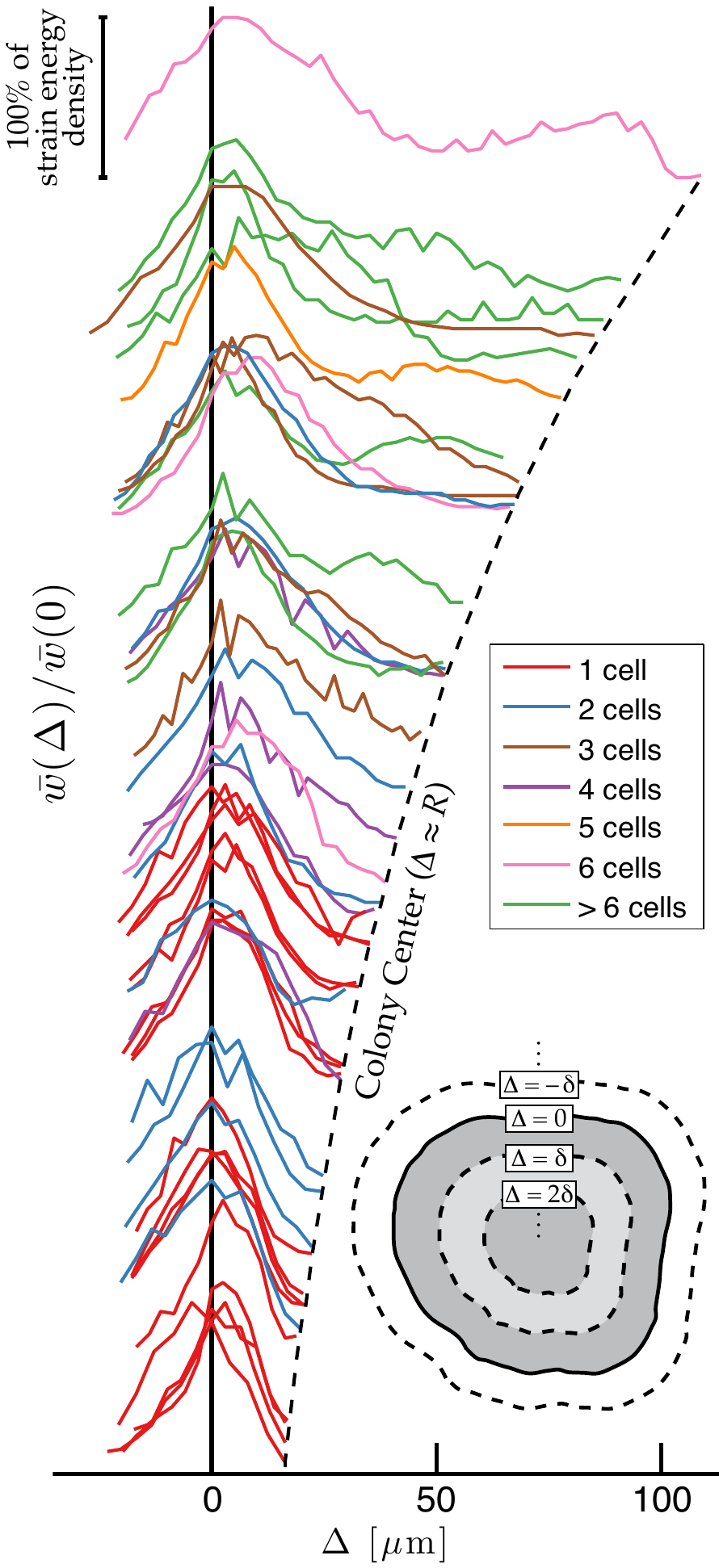}
\caption{\label{fig:profiles} (color online) Spatial distribution of strain energy for colonies of different size.
Each solid curve represents a colony's average strain energy density as a function of distance from the edge of the colony, $\Delta$.
For clarity, the profiles are spaced vertically according to the size of the colony.
Each profile terminates at the point where the inward erosion of the outer edge covers the entire area of the colony, at $\Delta \approx R$.
The erosion proceeds in discrete steps of size $\delta$, as illustrated in the inset.}
\end{figure}

To explore these trends, we measured traction stresses of 45 cohesive colonies of 1--27 cells.
For each colony, we defined an equivalent radius, $R$, as the radius of a disk with the same area.
The equivalent radii ranged from $20$ to $200\unit{\mu m}$.
We calculated the average strain energy density as a function of distance, $\Delta$, from the colony edge (Fig.~\ref{fig:profiles} inset).
Figure~\ref{fig:profiles} shows the normalized strain energy profiles, $\bar{w}(\Delta)/\bar{w}(0)$, of all 45 colonies. 
Usually, the strain energy density was largest near the colony edge ($\Delta=0$).
Because of the finite spatial resolution of our implementation of TFM, we measured some strain energy outside colony boundaries ($\Delta<0$).

\begin{figure}[hbt]
\includegraphics[width=.48\textwidth]{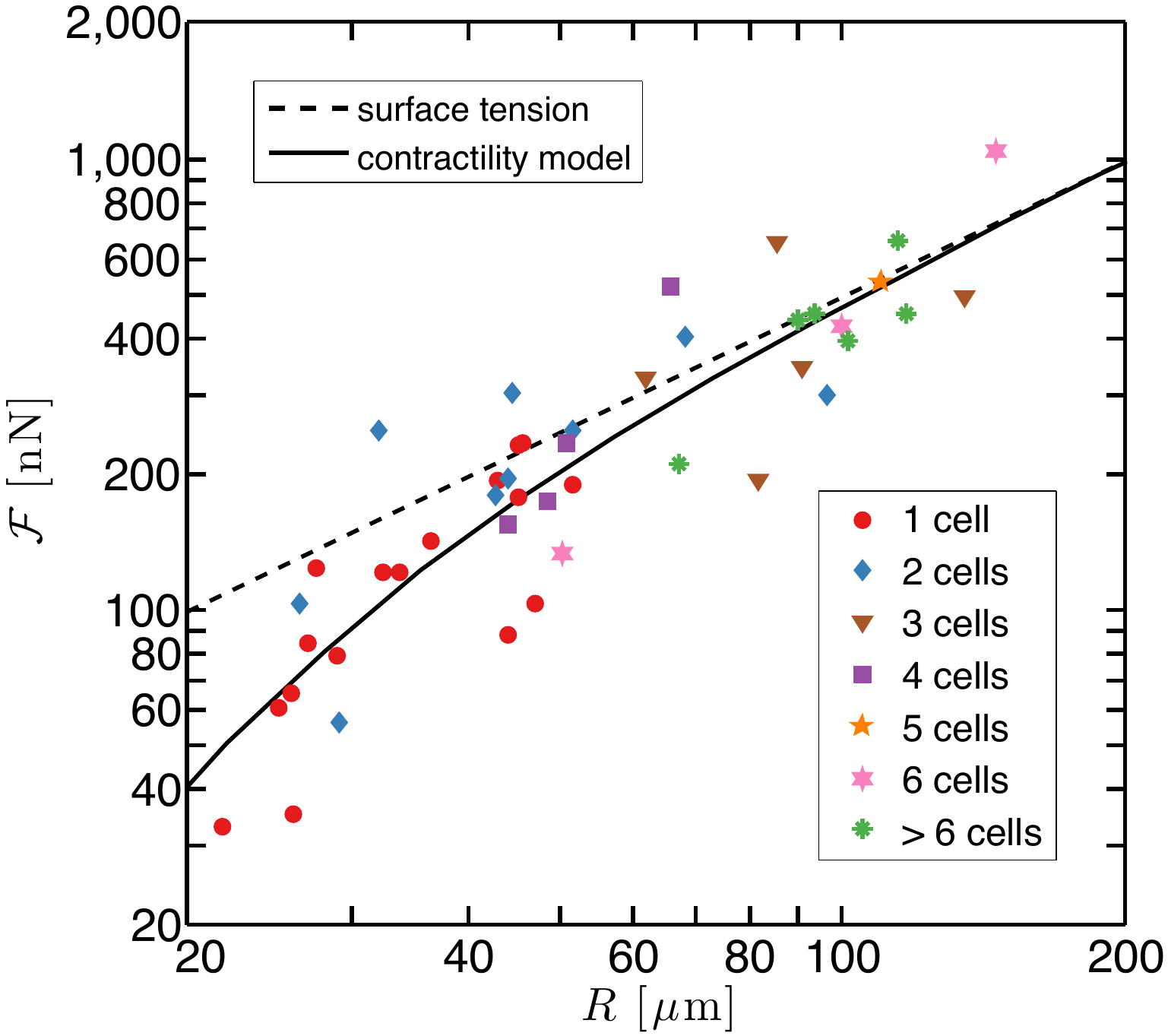}
\caption{\label{fig:scaling} (color online) Mechanical output of keratinocyte colonies versus geometrical size.
Total force transmitted to the substrate by the cell colonies, defined in Eq.~(\ref{eq:f}), is plotted as a function of the equivalent radius of the colonies.
The dashed line represents the scaling expected for surface tension, ${\cal F} \sim R$.
The solid line shows a fit of the data to the minimal contractility model in Eq.~(\ref{eq:fmodel}).}
\end{figure}

Next, we examined how global mechanical activity changes with the cell number and geometrical size of the colony.
As in previous studies, we calculated the ``total traction force" \cite{califano-2010, fu-2010},
\begin{equation}
\label{eq:f}
{\cal F}=\int dA \sqrt{\left(\sigma^s_{xz}\right)^{2}+\left(\sigma^s_{yz}\right)^2},
\end{equation}
exerted by the cell colony onto the substrate.
This quantity is meaningful when  stresses have radial symmetry and are localized at the colony edge, which is the case for the majority of colonies in this study.
We observed a strong positive correlation between equivalent radius and total force over the range of colonies examined (Fig.~\ref{fig:scaling}).
Similar trends have been seen for isolated cells over a smaller dynamic range of sizes  \cite{tolic-norrelykke-2005, califano-2010, tan-2003, fu-2010}.
We see no systematic differences in $\cal{F}$ for colonies of the same size having different numbers of cells, suggesting that cohesive cells cooperate to create a mechanically coherent unit.

The data in Fig.~\ref{fig:scaling}, while scattered, show clear monotonic growth of the mechanical output of cell colonies with their geometrical size, independent of the number of cells. 
%For smaller colonies ($R<60 \unit{\mu m}$), we cannot unambiguously determine the scaling exponent, which %could be between 2 and 3.
For smaller colonies  ($R<60 \unit{\mu m}$), the increase of total force is superlinear.
As the cell colonies get larger, the scaling exponent appears to approach unity.
We hypothesize that the transition to an apparently consistent exponent for the large colonies reflects the emergence of a scale-free material property of an adherent tissue, defined by the ratio ${\cal F}/(2\pi R) = (8 \pm 2) \times 10^{-4} \unit{N/m}$, with dimensions of surface tension.

Just as intermolecular forces yield the condensation of molecules into a dense phase, cohesive interactions between cells, mediated by cadherins, cause them to form dense colonies~\cite{gennes-2004,foty-2005}.
For large ensembles of molecules, molecular cohesion creates a free energy penalty per unit area, known as surface tension, for creating an interface between two phases.
It is tempting to think of the adherent colonies studied here as aggregates of cohesive cells that have wet the surface~\cite{douezan-2011}.
Indeed, when matter wets a surface, the traction stresses are localized at the contact line \cite{jerison-2011}, as we found in our cell colonies (Figs.~\ref{fig:TFM} and~\ref{fig:profiles}).

Effective surface tension of cell agglomerates has been invoked to explain cell sorting and embryogenesis \cite{steinberg-2007}. 
Previous measurements of non-adherent aggregates of cohesive cells reported effective surface tensions between $2$ and $20 \unit{mN/m}$ \cite{foty-1994, foty-1996, guevorkian-2010}.  
However, the origins of the effective surface tension of cohesive cells are distinct from conventional surface tension.
Recently, it was suggested that the surface tension is not only determined by contributions from cell--cell adhesions but also the contraction of acto-myosin networks \cite{kafer-2007,manning-2010}.
It is important to distinguish the effective surface tension due to active processes from the familiar surface tension defined in thermodynamic equilibrium.

To elucidate the origins of an effective surface tension in these experiments, we consider a minimal model proposed recently to describe cell--substrate interactions ~\cite{banerjee-2011a, edwards-2011}.
We describe a cohesive colony as an active elastic disk of thickness $h$ and radius $R$ (Fig.~\ref{fig:TFM}A).
The mechanical properties of the cell colony are assumed to be homogeneous and isotropic with Young's modulus $E$ and Poisson's ratio $\nu$.
Acto-myosin contractility is modeled as a contribution to the local pressure, linearly proportional to the chemical potential difference, $\Delta\mu$, between ATP and its hydrolysis products ~\cite{kruse-2005}.
In our model, the strength of cell--cell adhesions is implicitly contained in the material parameters of the colony, $E$ and $\nu$.
The constitutive equations for the stress tensor, $\sigma_{ij}$, of the colony are then given by
\begin{equation}
\label{eq:const}
\sigma_{ij}=\frac{E}{2(1+\nu)}\left[\frac{2\nu}{1-2\nu}\bm\nabla\cdot{\bf u} +\partial_iu_j+\partial_ju_i\right]+\delta_{ij}\zeta\Delta\mu,
\end{equation}
where ${\bf u}$ is the displacement field of the cell colony and $\zeta>0$ a material parameter that controls the strength of the active pressure, $\zeta\Delta\mu$.
Mechanical equilibrium requires that $\partial_j\sigma_{ij}=0$.

We use cylindrical coordinates and assume in-plane rotational symmetry.
The top surface is stress-free, $\sigma_{rz}|_{z=h+h_s}=0$, and we employ a simplified coupling of the colony to the substrate.
Ignoring all nonlocal effects arising from the substrate elasticity, $\sigma_{rz}|_{z=h_s}=Y u_r(z=h_s)\approx Y \bar{u}_r$.
Here, $u_r$ is the radial component of the displacement field, the bar denotes $z$-averaged quantities, and the rigidity parameter, $Y$, describes the coupling of the contractile elements of the colony to the substrate.
The local proportionality of stress and displacement is accurate only when the substrate thickness is much smaller than the characteristic length scale of the stress distribution or when the cells are on substrates of soft posts \cite{tan-2003}.

With these assumptions, the equation of force-balance simplifies to
\begin{equation}
\label{eq:eq}
\left[\partial_r(r\bar{ \sigma}_{rr} )-\bar{\sigma}_{\theta \theta}\right]/r =Y \bar{u}_r/h.
\end{equation}
Combining Eqs.~(\ref{eq:const}) and (\ref{eq:eq}), we find the governing equation for the radial displacement, $u_r$:
\begin{equation}
\label{eq:ur}
r^2\partial_r^2u_r + r\partial_r u_r - \left(1+r^2/\ell_p^{2}\right)u_r=0,
\end{equation}
where the penetration length, $\ell_p$, describing the localization of stresses near the boundary of the cell colony, is given by $\ell_p^2=E(1-\nu)h/\left[Y(1+\nu)(1-2\nu)\right]$.

The solution of Eq.~\eqref{eq:ur} with boundary conditions $u_r(r=0)=0$ and $\sigma_{rr}(r=R)=0$ can be expressed in terms of modified Bessel functions as
\begin{equation}
\label{eq:disp}
u(r)=-\zeta\Delta\mu \left[\frac{(1+\nu)(1-2\nu)}{E(1-\nu)} \right] R I_1(\beta r/R) A(\beta),
\end{equation}
with $\beta=R/\ell_p$ and $\left[A(\beta)\right]^{-1}=\beta I_0(\beta) - \left( \frac{1-2\nu}{1-\nu} \right) I_1(\beta)$.

As in our experiments, the resulting displacements and traction stresses are localized near the colony edge (Fig.~\ref{fig:profiles}).
To compare quantitatively to experiments, we calculate the total force,
\begin{equation}
\label{eq:fmodel}
{\cal F}(R)=2\pi Y \left| \int_0^R r dr\ u_r(r)\right|.
\end{equation}
In the large-colony limit, $R \gg \ell_p$, we find ${\cal F}(R) \simeq 2\pi \zeta\Delta\mu h R\sim R$, yielding the anticipated linear growth of total force for large colonies.
In this limit, the contractile active pressure dominates over internal elastic stresses and underlies the observed apparent surface tension.

The theory matches the scaling of the data reasonably well with $\ell_p=11\unit{\mu m}$ and apparent surface tension $\zeta\Delta\mu h \approx 8 \times 10^{-4} \unit{N/m}$, as shown by the solid line in Fig.~\ref{fig:scaling}. 
The penetration length, $\ell_p$, is comparable to the spatial resolution of our measurements.
For single cells, recent measurements have suggested apparent surface tensions of $2 \times 10^{-3}\unit{N/m}$ in an endothelial cell \cite{bischofs-2009} and $1 \times 10^{-4}\unit{N/m}$ in \emph{Dictyostelium} cells \cite{delanoe-ayari-2010}.
From previously published data on a millimeter-scale adherent sheet of cohesive cells, we calculated the apparent surface tension by integrating the average stress profiles near the sheet edge and found a value of about $7 \times 10^{-4}\unit{N/m}$ \cite{trepat-2009}.
For our cell colonies of thickness $h \approx 0.2 \unit{\mu m}$, estimated from confocal imaging of phalloidin-stained colonies, the fitted value of the apparent surface tension implies $\zeta\Delta\mu \approx 4 \unit{kPa}$.
This value is consistent with that inferred from experiments in crawling keratocytes~\cite{kruse-2006}. 
We can estimate the active pressure by assuming $\zeta\Delta\mu \approx \rho_m k_m \Delta_m$, where $\rho_m$ is the areal density of bound myosin motors, $k_m$ the stiffness of motor filaments, and $\Delta_m$ their average stretch.
Using $k_m \approx  1\unit{pN/nm}$, $\Delta_m \approx 1\unit{nm}$, and $\rho_m \approx 10^3 \unit{\mu m^{-2}}$, we find $\zeta\Delta\mu \approx 1 \unit{kPa}$~\cite{gunther-2007, banerjee-2011b}. 

In conclusion, we demonstrate a scaling relation between total traction force and the geometrical size of cohesive cell colonies adherent to soft substrates.
A simple physical model of cohesive colonies as adherent contractile disks captures the essential observations and suggests that the apparent surface tension in the large-colony limit is driven by acto-myosin contractility.
It is intriguing that a model of a cell colony with homogenous and isotropic properties is successful when the morphology of the underlying acto-myosin networks within the colony are patently heterogeneous and  anisotropic (Fig.~\ref{fig:TFM}).
Experiments measuring the apparent surface tension of colonies on substrates with different stiffnesses and with molecular perturbations that affect the contractility of acto-myosin networks and strength of intercellular adhesions will help to illuminate the limitations of the current model.
Additionally, the relationship between the apparent surface tension measured here in two-dimensional cell colonies and the effective surface tension measured in three-dimensional cell aggregates \cite{foty-1994, guevorkian-2010} needs to be established.
From a cell-biology perspective, it will be essential to determine the molecular mechanisms that regulate a colony's apparent surface tension.

\acknowledgements{We are grateful to Paul Forscher and Margaret L.\ Gardel for helpful discussions and to an anonymous reviewer for useful comments.
This work was supported by NSF Graduate Research Fellowships to A.F.M. and C.H., NSF grants to E.R.D.\ (DBI-0619674) and M.C.M.\ (DMR-0806511 and DMR-1004789), NIH grants to V.H.\ (AR054775 and AR060295), as well as support to E.R.D.\ from Unilever.}

%\bibliography{colony}

\end{document}